       \documentclass{revtex4}
       \usepackage{epsfig}

\begin{document}
       \title{Comparison of gain-loss asymmetry behavior for stocks and indexes}
       \author{Magdalena Za{\l}uska-Kotur}
       \address{ Institute of Physics, Polish Academy of Sciences,
         Aleja Lotnik\'ow 32/46,
         02--668 Warsaw, Poland}
       \author{Krzysztof Karpio}
       \address{Department of Econometrics and Informatics, Warsaw Agricultural University, ul Nowoursynowska 166, 02-787 Warsaw, Poland,
       \\ Institute for Nuclear Studies, ul. Ho\.{z}a 69, 00-681 Warsaw, Poland}
       \author{Arkadiusz Or{\l}owski}
       \address{Department of Econometrics and Informatics, Warsaw Agricultural University, ul Nowoursynowska 166,
       02-787 Warsaw, Poland\\
       Institute of Physics, Polish Academy of Sciences, Aleja Lotnik\'ow 32/46, 02--668 Warsaw, Poland }

       \begin{abstract}
       Investment horizon approach has been used to analyze indexes of Polish stock market.
       Optimal time horizon for each return value is evaluated by fitting appropriate function form of the distribution.
       Strong asymmetry of gain-loss curves is observed for WIG index, whereas gain and loss curves look
       similar for WIG20 and for most of individual companies stocks. The gain-loss asymmetry for these data,
       measured by the coefficient, that we postulated before \cite{karpio}, has opposite sign to this for WIG index.
       \end{abstract}

       \maketitle
       Statistical analysis of market indexes becomes source of detailed knowledge about character and relations
       between economical processes \cite{mantegna,hui,bouchard}.
       Recently invented investment horizon analysis \cite{simonsen-ex,simonsen-tur,simonsen2,simonsen3,simonsen4}
       is an approach based on the inverse statistics. This type of analysis deals with the distribution
       of time distances between chosen moment and moment when given return value is obtained for the first time.
       Such time distance measured in the random walk problem is called the first passage time.
       The statistics of the first passage time for the classical random walk is given by the distribution
       \begin{equation}
       \label{random_walk}
       p(t)=a \frac{\exp(-\frac{a^2}{t})}{\sqrt{\pi} t^{3/2}},
       \end{equation}
       where $a$ is a distance that we want to reach. When we treat asset prices $S(t)$ as the random process,
       returns at time $\Delta t$ are measured as $\ln S(t)-\ln S(t-\Delta t)$. Hence to get
       distribution
       of times for given return value we will use variable $s(t)=\ln S(t)$.
       After taking a logarithm of data we subtract trend $d(t)$ of the data, thus getting data $s \tilde = s-d(t)$.
       The mean trend is calculated as moving average over 100 points.
       \begin{figure}
       \includegraphics[width=8.5cm,angle=-90]{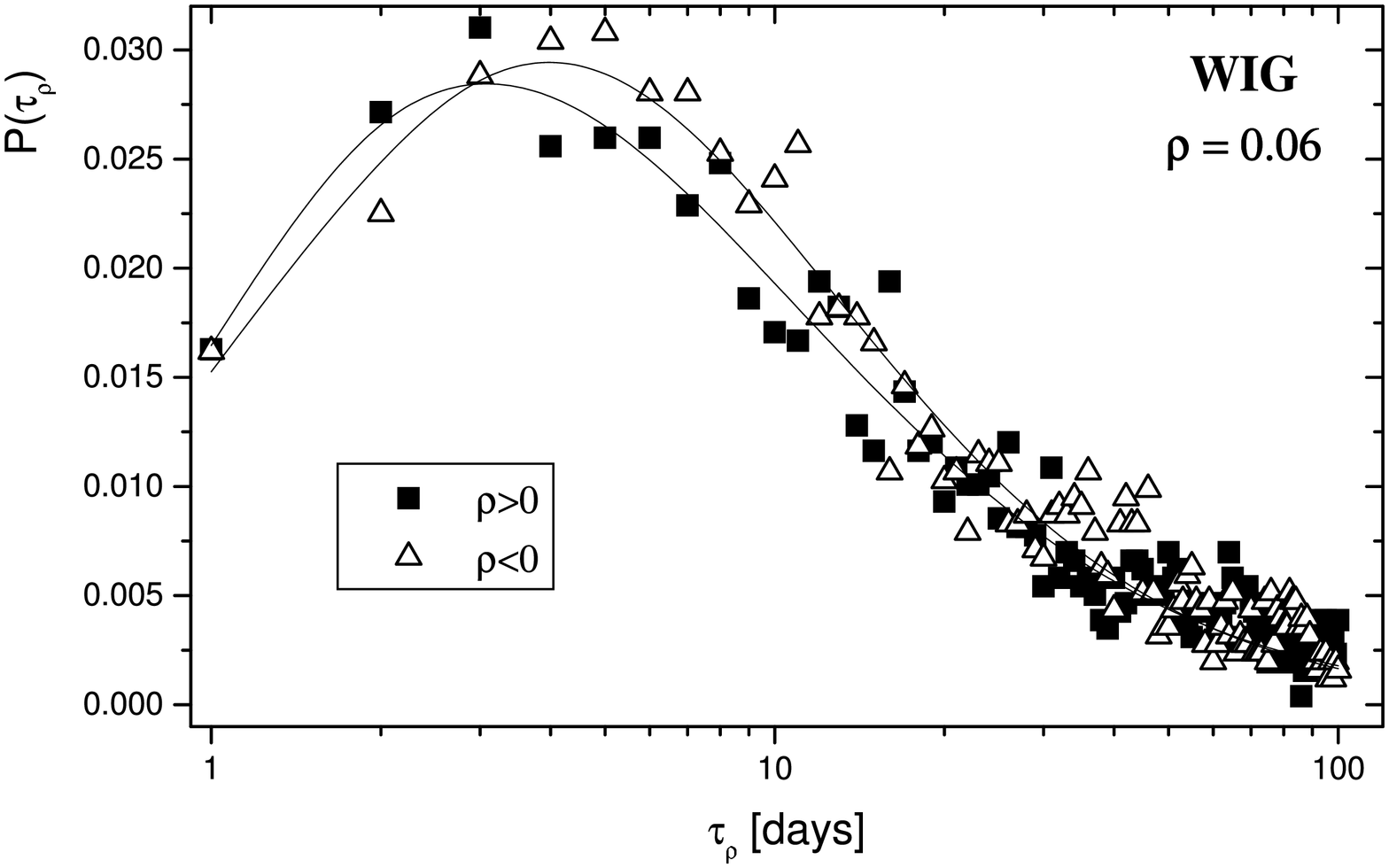}
       \caption{\label{fig:2}WIG investment horizon distribution calculated for return values
       $\rho=0.06$ - closed squares and $\rho=-0.06$ - open triangles. }
       \end{figure}
       \begin{figure}
       \includegraphics[width=8.5cm,angle=-90]{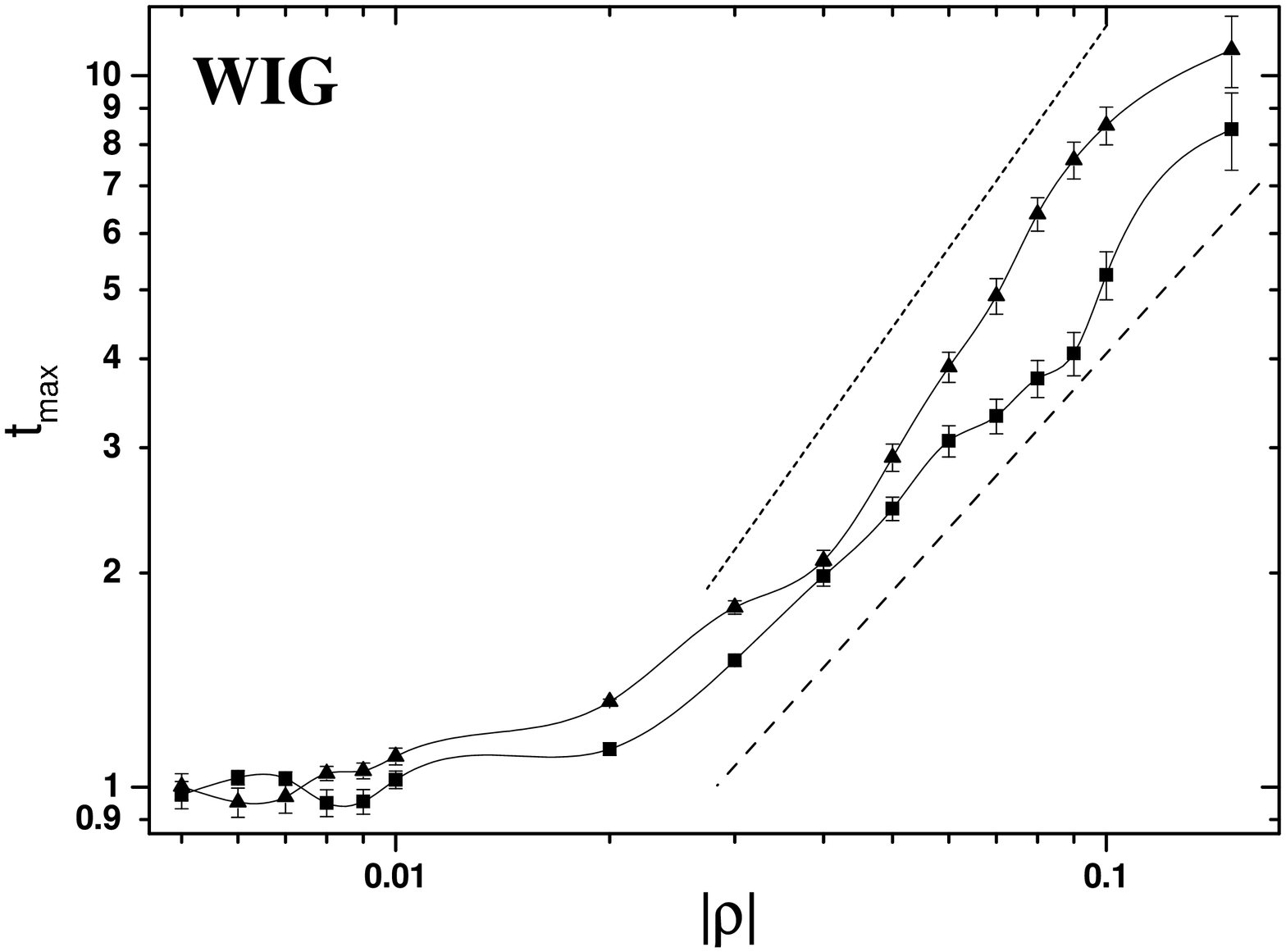}
       \caption{\label{fig:2} Optimal investment horizon plotted as a function of absolute return value for WIG.
       Data for $\rho>0$ are marked by squares, and for $\rho<0$ by triangles. Dashed lines show average slope
       of gain and loss curves.}
       \end{figure}

       In the inverse statistics we begin with histograms of investment horizons, that are build up by starting
       with at different moments in the index history, and measuring time that is needed to obtain assumed return
       value for the first time. We have one histogram for each return value.
       When we try to describe investment horizon distribution by (\ref{random_walk}) it appears that the fit is poor.
       The low time branch has too small values. Other type distribution has been postulated \cite{simonsen3}
       \begin{equation}
       \label{D}
       p(t)=\frac{\nu}{\Gamma(\frac{\alpha-1}{\nu})} \frac{\beta^{2(\alpha-1)}}{(t-t_0)^{\alpha}}\exp\{-(\frac{\beta^2}
       {t-t_0})^\nu\},
       \end{equation}
       where $\alpha, \beta, \nu$ and $t_0$ are parameters of this distribution. Now, both (\ref{random_walk})
       and (\ref{D}) decay for large $t$ as
       \begin{equation}
       p(t) \approx t^{\alpha},
       \end{equation}
       which has been checked for data from different markets \cite{simonsen3,zhou}. From (\ref{D}) we can evaluate the value $t$ of maximal probability
       \begin{equation}
       t_{max}=t_0+\beta^2 (\frac{\nu}{\alpha+1})^{1/\nu},
       \end{equation}
       which actually has been used as fitting parameter.

       Summing up the procedure we follow for each studied data we have: find time horizon distribution for given return value, fit (\ref{D}) and obtain $t_{max}$.  We repeat the procedure for each return value separately, and finally we plot $t_{max}$ vs $\rho$ for positive and for negative values of $\rho$.
       If our process was ideal random process two curves plotted in such a way would lie on a one curve, with slope equal to $2.0$. As it has been shown before for DIJA that is not true \cite{simonsen-tur,simonsen2}. Slope of curves is actually smaller, as it has been shown, moreover there is an asymmetry: gain curve goes above loss one \cite{simonsen2,simonsen4}.
       \begin{figure}[h]
       \includegraphics[width=8.5cm,angle=-90]{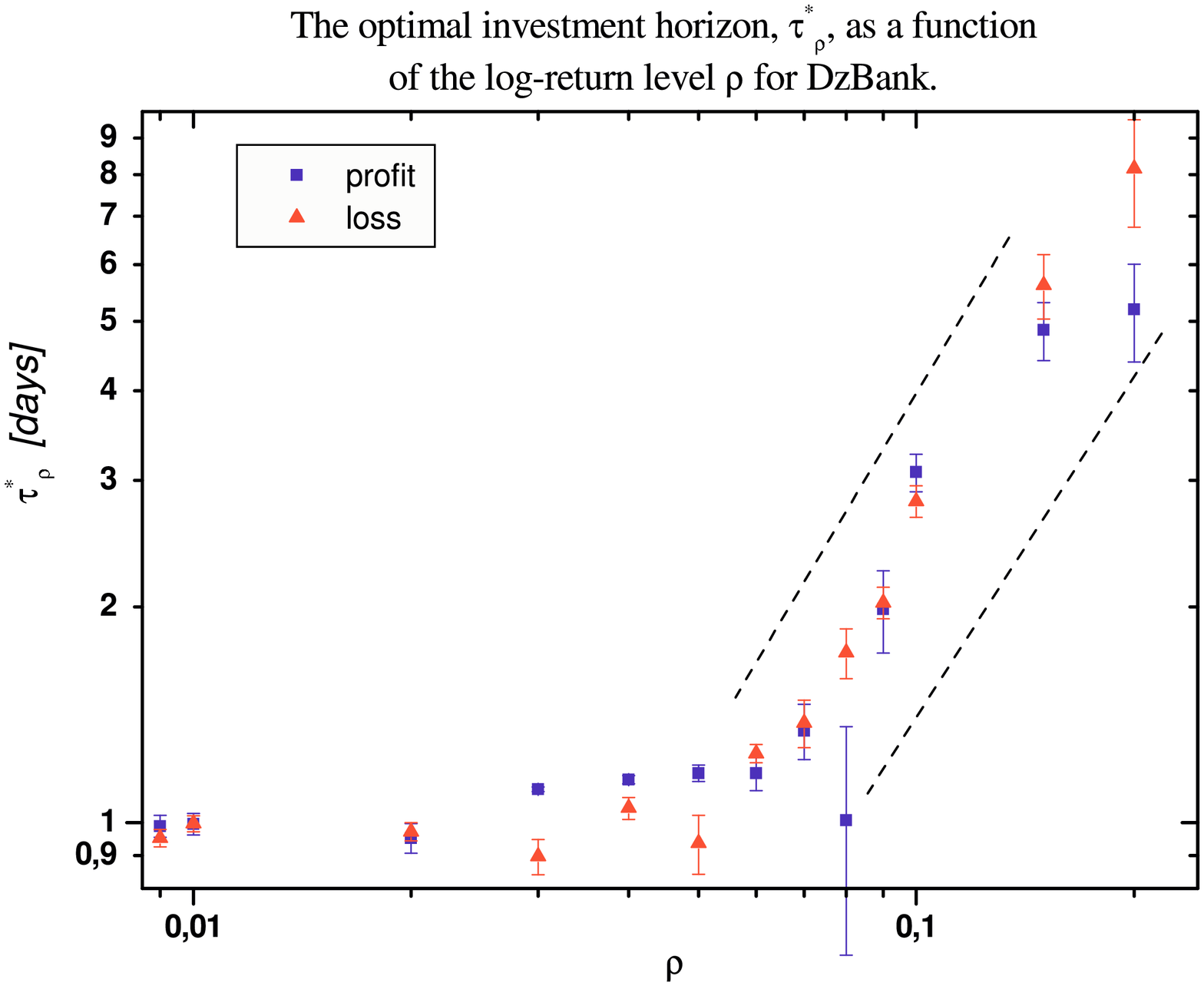}
       \includegraphics[width=8.8cm,angle=-90]{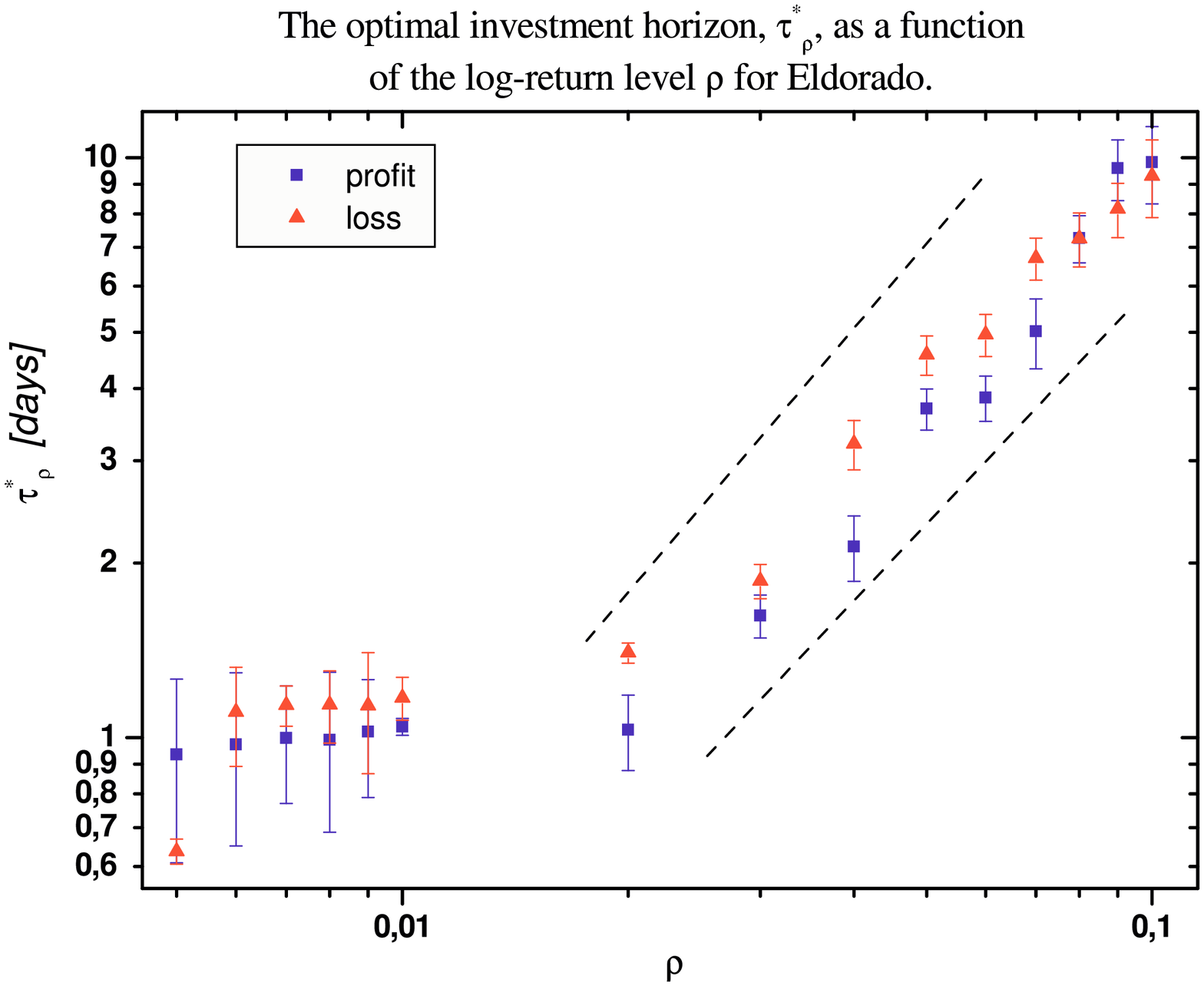}
       \includegraphics[width=8.5cm,angle=-90]{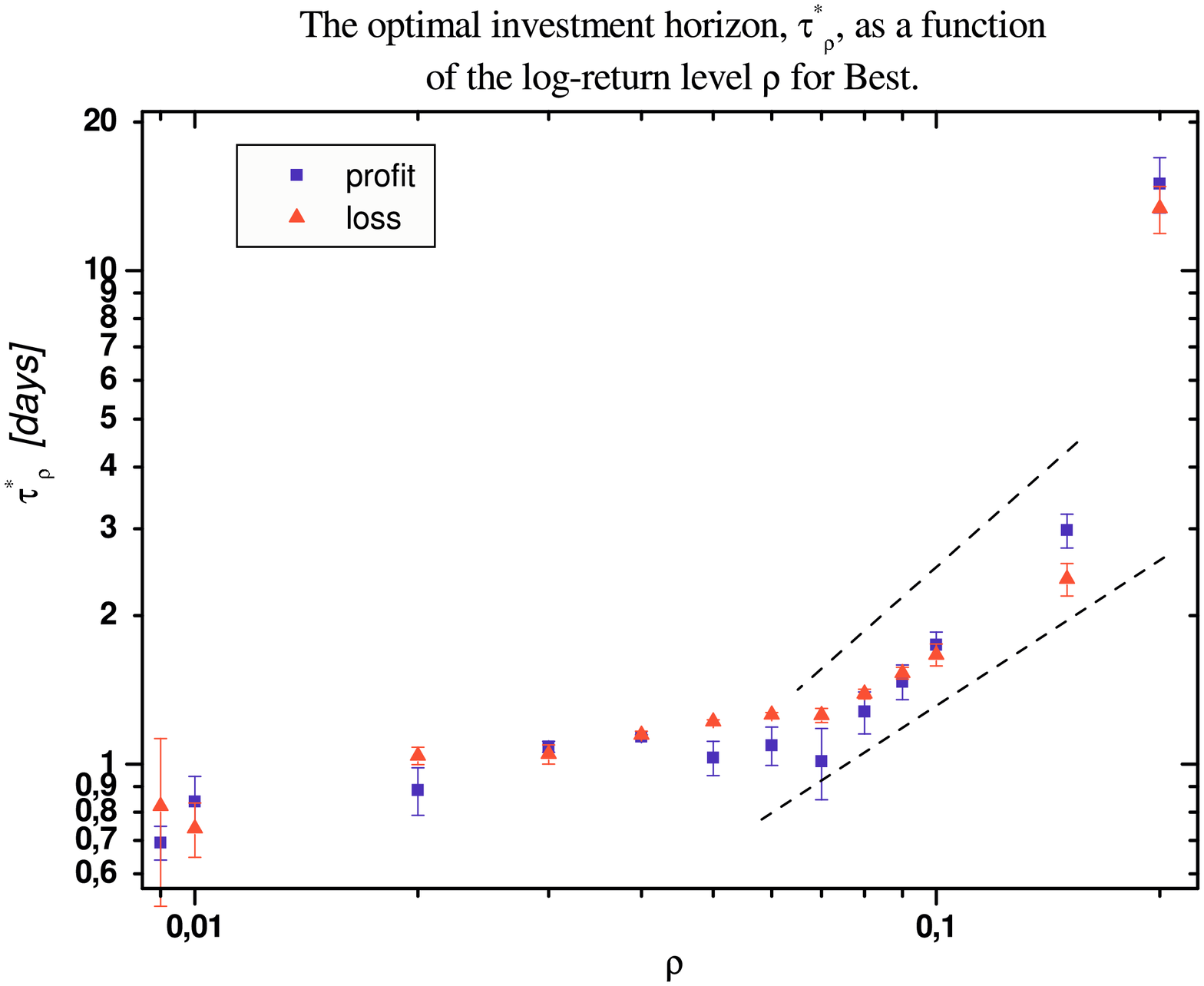}
       \includegraphics[width=8.5cm,angle=-90]{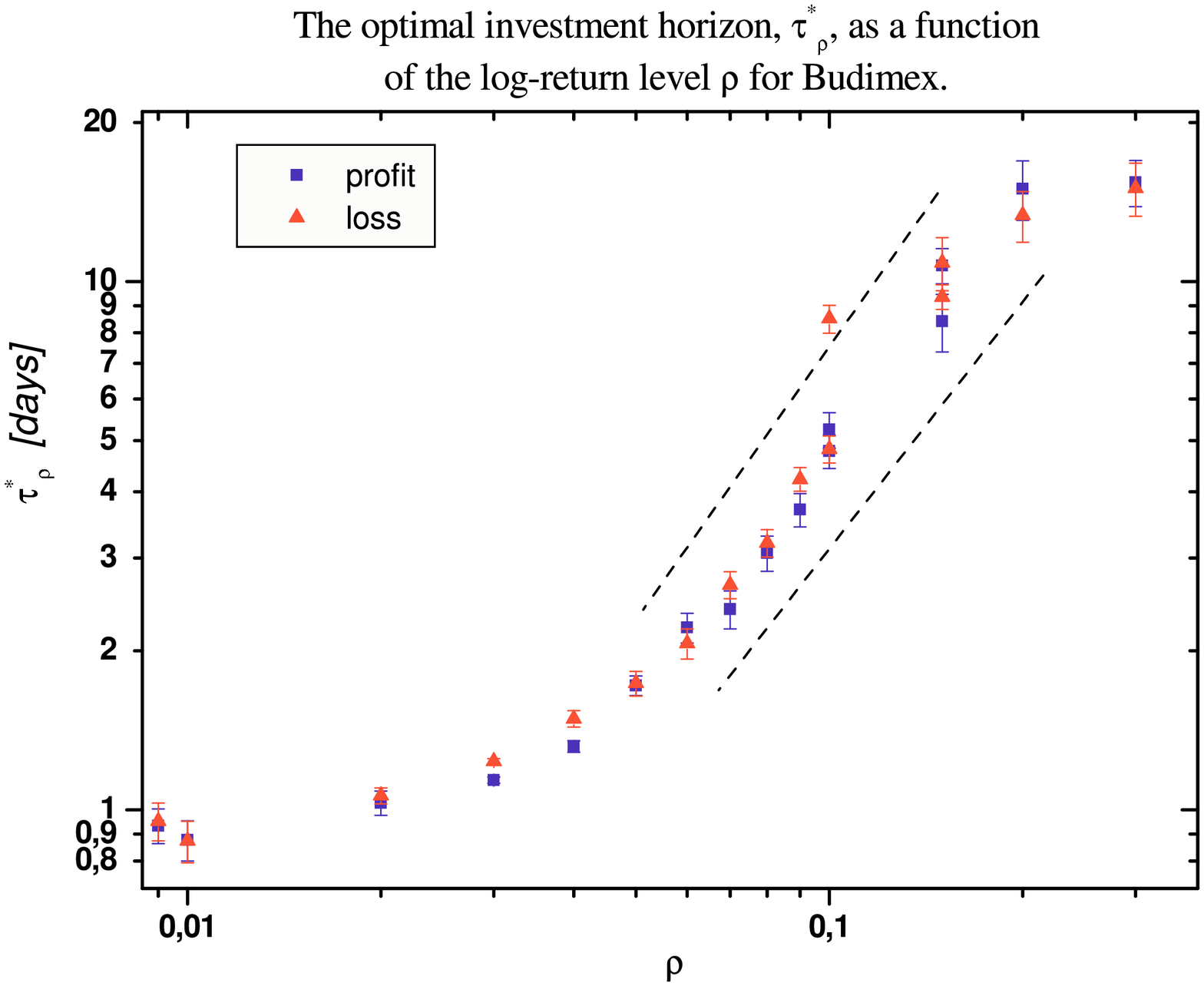}
\caption{\label{fig:2} Optimal investment horizon plotted as a function of absolute return value for a) Best, b) Budimex, c) DzBank and d) Eldorado. Data for $\rho>0$ are marked by squares, and for $\rho<0$ by triangles. Dashed lines show average slope of gain and loss curves.}
       \end{figure}


       We have shown, that the asymmetry of histograms for gain-loss return values is also present when
       data for Polish index WIG are analyzed \cite{karpio}. An example of such distribution is shown in Fig 1.
       There is however one basic difference between results for WIG and for DJIA. Asymmetry of WIG is opposite
       to that noticed in the case of DJIA. In the first case histogram for gain data is closer to the axis,
       and has maximum at lower value than histogram for loss data. Such situation is typical for all analyzed
       return values, which can be seen in Fig 2, where $t_{max}$ vs $|\rho|$ plot is shown.
       We have analyzed indexes for several East European market \cite{karpio}, and found the same behavior
       for all of them. At the same time Austrian ATX follows tendency observed for DJIA. It seems, that
       ``general rule'' saying that it is much more difficult to gain that to lose money applies only to developed
       markets, whereas for emerging markets the opposite is true.
       Is it also true for stocks of individual companies that are components of WIG?
       In other words the question is whether WIG behavior is a simple sum of individual stocks behavior.
       We calculated gain-loss curves for several main companies. Some results are shown in Fig. 3
       We can see that both gain and loss curves lie at the almost same curve. It means, that index WIG
       does not behave as a simple sum of individual stocks. Let us analyze our results further and find
       linear approximations of gain and loss curves. In Ref. \cite{karpio} we defined the asymmetry
       measure $\kappa=\gamma-\gamma^{'}$, where $\gamma$ and $\gamma ^{\prime}$ are slopes of fits to gain
       and loss curves respectively. For all studied East European indexes parameter $\kappa$ was negative.
       For stocks of companies above we have $\kappa=0.42,0.1,0$ and $0.3$ in the same order as above:
       Best, Budimex, DzBank, and Eldorado. Thus not only we cannot see any significant difference between
       gain and loss curves, but also their slope are in the reverse order, when compared to WIG. Stocks
       for other companies behave in similar way. Moreover curves plotted for index WIG20 in Fig 4 look
       like simple sum of these for individual companies with $\kappa=0.1$.
       \begin{figure}[h]
       \includegraphics[width=8.5cm,angle=-90]{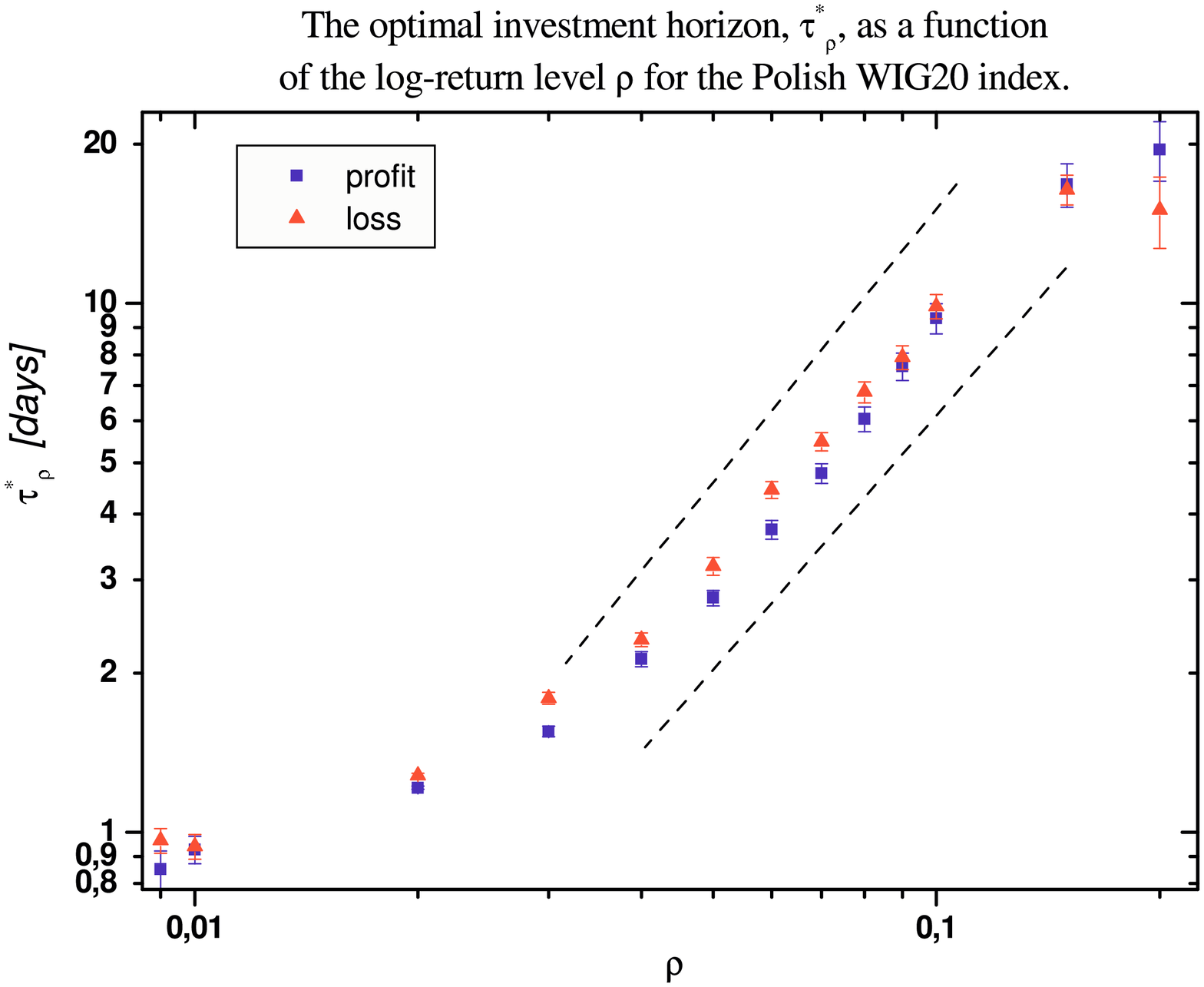}
       \caption{\label{fig:2} Optimal investment horizon plotted as a function of absolute return value for WIG20.
       Data for $\rho>0$ are marked by squares, and for $\rho<0$ by triangles. Dashed lines show average slope of
       gain and loss curves.}
       \end{figure}

       Hence what is the reason, that WIG is so different? It is calculated in different way,
       it contains more companies, and that can lead to correlated behavior of its elements.
       Such possible correlations would explain strange time vs return dependence of gain-loss curves
       for the emerging market indexes. Is the above explanation convincing
       enough? This and related problems will be undertaken in a
       forthcoming paper.


\begin{thebibliography}{00}
       \bibitem{karpio} K. Karpio, M. A. Za³uska-Kotur, A. Or{\l}owski, submitted to Physica
       A.
       \bibitem{mantegna} R.N. Mantegna, H. E. Stanley,
       An introduction to Econophysics: Correlations and Complexity in Finance, CUP, Cambridge, England,
       2000.
       \bibitem{hui}N F Johnson, P Jefferies, and P M Hui, Financial Market Complexity, Oxford University Press,
       2003.
       \bibitem{bouchard} J P Bouchard and M Potters, Theory of Financial Risks:
       from Statistical Physics to Risk Management, Cambridge University Press,
       2000.
       \bibitem{simonsen-ex}
       M.H. Jensen, A. Johansen, F. Petroni, and I. Simonsen, Physica A {\bf 340}, 678
       (2004).
       \bibitem{simonsen-tur}
       M.H. Jensen, A. Johansen, and I. Simonsen,
       Int. J. Mod. Phys. B {\bf 17}, 4003 (2003).
       \bibitem{simonsen2}
       M.H. Jensen, A. Johansen, and I. Simonsen, Physica A {\bf 324}, 338 (2003).
       \bibitem{simonsen3}
       I. Simonsen, M.H. Jensen, and A. Johansen, Europ. Phys. J B {\bf 27}, 583 (2002).
       \bibitem{simonsen4} M.H. Jensen, A. Johansen, I. Simonsen, and F. Petroni, Physica A {\bf 340}, 678(2004).
       %
       \bibitem{zhou}W.X. Zhou, W.K. Yuan, Physica A {\bf 353}, 433 (2005).
       \end{thebibliography}
       \end{document}